\RequirePackage{lineno}
\documentclass[preprint,amsmath,amssymb,superscriptaddress,unsortedaddress]{revtex4}

\usepackage{epsf}

\usepackage{color}
\usepackage[dvipdfmx]{hyperref}
\usepackage{graphicx}
\usepackage[normalem]{ulem}
\usepackage{newfloat}
\usepackage[latin1]{inputenc}
\usepackage{comment}
\usepackage{amsmath}
\usepackage{bm}
\usepackage[version=3]{mhchem}

\newcommand{\m}{\textmu}

\begin{document}
	
	\title{Laser-based angle-resolved photoemission spectroscopy with micrometer spatial resolution and detection of three-dimensional spin vector}
	\author{Takuma Iwata}
	\affiliation{Graduate School of Advanced Science and Engineering, Hiroshima University, 1-3-1 Kagamiyama, Higashi-Hiroshima 739-8526, Japan}
	\affiliation{International Institute for Sustainability with Knotted Chiral Meta Matter (WPI-SKCM$^2$), Hiroshima University, Higashi-hiroshima 739-8526, Japan}
	
	\author{T.~Kousa} 
	\affiliation{Graduate School of Advanced Science and Engineering, Hiroshima University, 1-3-1 Kagamiyama, Higashi-Hiroshima 739-8526, Japan}
	
	\author{Y.~Nishioka} 
	\affiliation{Graduate School of Advanced Science and Engineering, Hiroshima University, 1-3-1 Kagamiyama, Higashi-Hiroshima 739-8526, Japan}
	
	\author{K.~Ohwada} 
	\affiliation{Graduate School of Advanced Science and Engineering, Hiroshima University, 1-3-1 Kagamiyama, Higashi-Hiroshima 739-8526, Japan}

	\author{K.~Sumida} 
	\affiliation{Materials Sciences Research Center, Japan Atomic Energy Agency, Sayo-gun, Hyogo 679-5148, Japan}

	\author{E.~Annese}
	\affiliation{Brasilian Center for Physical Science, R. Dr. Xavier Sigaud, 150 - Urca, Rio de Janeiro - RJ, 22290-180, Brazil}
	
	\author{M.~Kakoki}
	\affiliation{Graduate School of Advanced Science and Engineering, Hiroshima University, 1-3-1 Kagamiyama, Higashi-Hiroshima 739-8526, Japan}

   \author{Kenta~Kuroda}
	\affiliation{Graduate School of Advanced Science and Engineering, Hiroshima University, 1-3-1 Kagamiyama, Higashi-Hiroshima 739-8526, Japan}
	\affiliation{International Institute for Sustainability with Knotted Chiral Meta Matter (WPI-SKCM$^2$), Hiroshima University, Higashi-hiroshima 739-8526, Japan}
	
	\author{H.~Iwasawa}
	\affiliation{Institute for Advanced Synchrotron Light Source, National Institutes for Quantum Science and Technology, Sendai 980-8579, Japan}
	\affiliation{Synchrotron Radiation Research Center, National Institutes for Quantum Science and Technology, Hyogo 679-5148, Japan}
	\affiliation{Hiroshima Synchrotron Radiation Center, Hiroshima University, 2-313 Kagamiyama, Higashi-Hiroshima 739-0046, Japan}

	\author{M.~Arita}
	\affiliation{Hiroshima Synchrotron Radiation Center, Hiroshima University, 2-313 Kagamiyama, Higashi-Hiroshima 739-0046, Japan}
	
	\author{S.~Kumar}
	\affiliation{Hiroshima Synchrotron Radiation Center, Hiroshima University, 2-313 Kagamiyama, Higashi-Hiroshima 739-0046, Japan}

	\author{A.~Kimura}
	\affiliation{Graduate School of Advanced Science and Engineering, Hiroshima University, 1-3-1 Kagamiyama, Higashi-Hiroshima 739-8526, Japan}
	\affiliation{International Institute for Sustainability with Knotted Chiral Meta Matter (WPI-SKCM$^2$), Hiroshima University, Higashi-hiroshima 739-8526, Japan}
		
	\author{K.~Miyamoto} 
	\affiliation{Hiroshima Synchrotron Radiation Center, Hiroshima University, 2-313 Kagamiyama, Higashi-Hiroshima 739-0046, Japan}

	\author{T.~Okuda}
	\affiliation{Hiroshima Synchrotron Radiation Center, Hiroshima University, 2-313 Kagamiyama, Higashi-Hiroshima 739-0046, Japan}

\begin{abstract}
We have developed a state-of-the-art apparatus for laser-based spin- and angle-resolved photoemission spectroscopy with micrometer spatial resolution ({\m}-SARPES).
This equipment is achieved through the combination of a high-resolution photoelectron spectrometer, a 6-eV laser with high photon flux that is focused down to a few micrometers, a high-precision sample stage control system, and a double very-low-energy-electron-diffraction spin detector.
The setup achieves an energy resolution of 1.5 (5.5) meV without (with) the spin detection mode, compatible with a spatial resolution better than 10 \m m. This enables us to probe both spatially-resolved electronic structures and vector information of spin polarization in three dimensions. 
The performance of {\m}-SARPES apparatus is demonstrated by presenting ARPES and SARPES results from topological insulators and Au photolithography patterns on a Si (001) substrate.
\end{abstract}

\date{\today}
 \maketitle
\section{Introduction}
Understanding the electron behaviors in solids has become increasingly important because it is closely connected not only to fascinating quantum properties of condensed matters but also to their numerous functionalities relevant to device applications~\cite{Ashcroft1976-qb}.
The fundamental properties of the electrons are predominantly described by distinct quantum parameters including energy, momentum, and spin. In this context, angle-resolved photoemission spectroscopy (ARPES)~\cite{Damascelli2003-lb, Hufner_2007} is widely known as one of the leading experimental techniques in the research field~\cite{Lv2019,Sobota2021-dj,Zhang2022, King2021chemrev}, since it can directly probe energy- and momentum-resolved electronic structures and Fermi surfaces. 

In the past decades, technological developments of ARPES have been achieved through effective combinations of characteristic light sources and state-of-the-art electron spectrometers. As a remarkable example, ultrahigh energy-resolution in the meV range has been successfully attained in ARPES through the utilization of ultraviolet lasers~\cite{Shimojima2015-hd, iwasawa2017elsevier,Tamai_prb2013}, showcasing its effectiveness in elucidating the low-energy excitation structures of electrons~\cite{Okazaki_Science2012,Huang2016,Tamai_prb2013} and many-body interactions~\cite{Kondo_NatureCom2015,Koralek_prl2006,Arai_natureMat2022}.
Recently, improvements in the spatial resolution of the ARPES technique have received a lot of attention~\cite{Rotenberg2014jsr,iwasawa2020iop, cattelan2018nanomaterials}.
In this technique, the excitation photon beam is focused onto the sample down into a very small spot, either in a $few$-$micron$ or $sub$-$micron$ scale, often referred to as \m-ARPES or nano-ARPES.
The capability of these techniques has successfully revealed local electronic structures, such as phase-separated electronic structures in strongly correlated materials~\cite{Watson_npj2019,kuroda2020naturecommn, bao2021commnphys}, two-dimensional flakes~\cite{Sakano2022PhysRevResearch}, and edge states of topological materials~\cite{noguchi2019nature,Yu_prx2021}.

All of the aforementioned ARPES techniques are widely used.
However, they are still unable to probe spin degree of freedom, despite its significant role in many quantum properties of materials, such as magnetism~\cite{fmmode_rmp2004}, superconductivity~\cite{Lee_rmp2006,Stewart_rmp2011}, Rashba spin-orbit coupling~\cite{Okuda2013-wg, Manchon_NatureMat2015}, band topology~\cite{RevModPhys.82.3045, Armitage_RMP2018}, and spin-valley~\cite{Xu_NaturePhys2014}.
The integration of spin detectors into ARPES photoelectron spectrometers has added spin-resolution capabilities to ARPES (SARPES)~\cite{okuda2017iop,Dil_iop2019,TUSCHE2015ultramicro}, allowing to access spin information of energy/momentum-resolved band structures.
Moreover, comprehensive information about spin polarization in three-dimensions can be obtained by employing multiple spin detectors~\cite{Hoesh_jrsrp2002, Okuda2015-ug}, which is indispensable for accurately determining the quantization axis and spin textures in quantum materials.
Although spin detection is highly time-consuming due to the relatively low efficiency of existing spin polarimeters~\cite{okuda2017iop}, light sources with high photon flux, such as 3rd generation synchrotrons~\cite{Hoesh_jrsrp2002, okuda2011revsciins} and lasers~\cite{Jozwiak_rsi2010,Gotlieb_rsi2013,yaji2016revsciins}, can compensate for this low efficiency.
Recently, SARPES has been achieved with a sub-micrometer spatial resolution using a laser light source~\cite{xu2023rsi}.
However, realizing SARPES simultaneously with high-resolution, spatial-resolution, and spin-vector-resolution is still challenging.

In this paper, we describe a setup of SARPES apparatus with a combination of multiple functions.
We overcome the low efficiency of spin detection measurements by utilizing very-low-energy-electron-diffraction (VLEED) spin detectors~\cite{okuda2017iop} and a 6-eV laser~\cite{Koralek_rsi2007,Jiang_rsi2014}.
In addition, this apparatus is equipped with double VLEED spin detectors, which permits a vector analysis of photoelectron spin-polarizations in three dimensions~\cite{Okuda2015-ug}. 
Thanks to the inherent characteristics of the laser light source, such as its high coherence and monochromaticity, we achieve a high energy resolution of 1.5~meV (5.5~meV) without (with) the spin detection mode, while maintaining a good spatial resolution better than 10 {\m}m.
Present {\m}-SARPES is able to give us a great capability for quickly mapping out not only spatially-resolved electronic band structures but also their spin textures.

\section{Laser \m-SARPES system}
\begin{figure*}[t]
\begin{center}
\includegraphics[width=\linewidth]{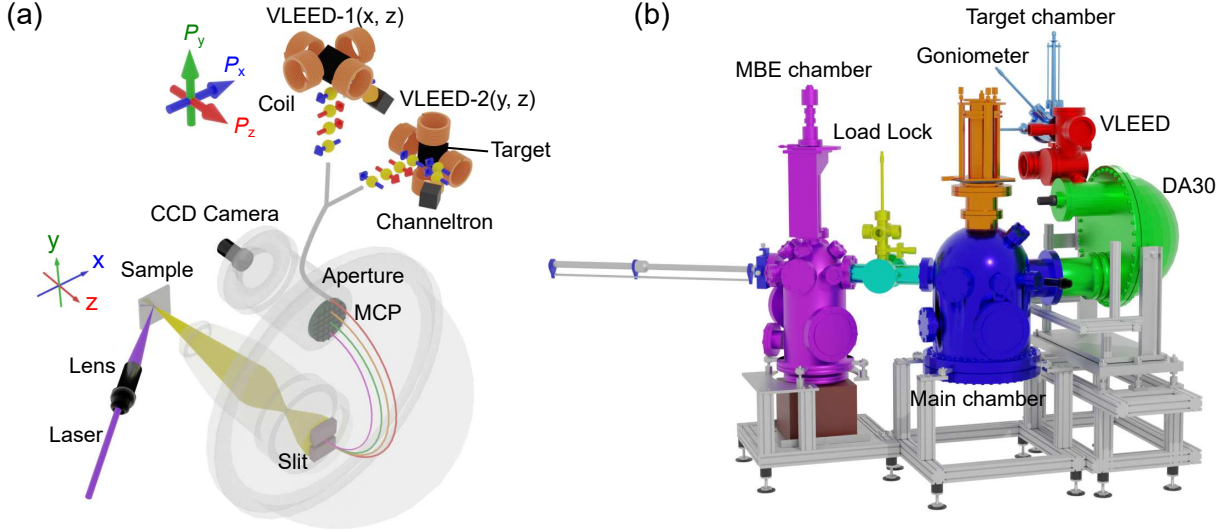}
\caption{(a) Schematic design of laser \m-SARPES with double VLEED spin detectors probe spin components ($P_x$, $P_y$, and $P_z$)~\cite{Hoesh_jrsrp2002, Okuda2015-ug}.
(b) Overview of the scaled layout of our laser \m-SARPES apparatus.}
\label{fig1}
\end{center}
\end{figure*}

Figure~\ref{fig1}(a) illustrates a schematic design of key parts of our \m-SARPES apparatus. An ultraviolet laser ($h\nu$=5.9-6.49~eV) generated via the 4th harmonics of picosecond infrared pulses from Ti-sapphire lasers is used for photoelectron excitations.
The laser beam is focused onto the sample surface at the micrometer level using a quartz lens assembly~\cite{iwasawa2017elsevier}.
The high photon flux of the laser~\cite{Gotlieb_rsi2013,yaji2016revsciins} can compensate for the low efficiency of spin detections~\cite{okuda2017iop} and thus shortens the acquisition time.
In addition, the 6-eV laser can penetrate through the atmosphere, and therefore it is much easy to handle optical parameters and optimize setup compared to other light sources (ex. 7-eV laser~\cite{Shimojima2015-hd} and 11-eV laser~\cite{He2016-yo,Peli2020-gm,Lee_rsi2020,kawaguchi_arxiv2023}) and ultraviolet synchrotron radiations.
  
The excited photoelectrons are collected by a hemispherical electron analyzer with a multi-channel plate (MCP) to image electronic band structures.
The photoelectrons passing through a size variable aperture above the MCP are transferred to two different VLEED spin-detectors (VLEED-1 and 2), and the reflected photoelectrons by Fe(001)$p$(1$\times$1)-O targets~\cite{Bertacco_prb1999,okuda_rsi2008} are detected by channeltrons.
The target in VLEED-1(2) can be magnetized by electric coils along $x$ and $z$ ($y$ and $z$) directions corresponding to the detection axes of the spin polarization $P_x$ and $P_z$ ($P_y$ and $P_z$).
Thus, thanks to double VLEED detectors~\cite{Okuda2015-ug}, one can fully determine a spin vector of the photoelectrons in three dimensions, which is very important for understanding the intrinsic properties of quantum materials.

Figure~\ref{fig1}(b) presents an overview of our SARPES apparatus, consisting of three ultrahigh vacuum (UHV) chambers: main, target preparation, and molecular beam epitaxy (MBE) chambers.
In the main chamber, the hemispherical electron analyzer, ScientaOmicron DA30L is equipped with double VLEED spin detectors.
The analyzer has an electron deflector function and collects the photoelectrons in a range of emission angles ($\theta_{x}$, $\theta_{y}$) without a sample rotation, which brings significant advantages in spatial resolution measurements. Here, $\theta_x$ and $\theta_y$ are defined as an emission angle along the horizontal and vertical axes of an analyzer slit, respectively.
To achieve a high spatial-resolution in \m-SARPES, we adopted a high-precision sample manipulator, which consists of a high-precision XYZ translator stage (iXYZ, the ExPP Co. Ltd.), a high-precision rotary stage (iRS152, Vacuum and Optical Instruments Co. Ltd.) \cite{aiura2012revsciins} and a goniometer with a liquid He cryostat ($i$-Gonio LT, R-Dec Co. Ltd.) \cite{aiura2003revsciins}.
These motor motions are controlled by a software coded using LabVIEW (National Instruments).

The details of our laser source and frequency conversion optics to generate 6-eV laser are described elsewhere~\cite{iwasawa2017elsevier}.
The assembly of the microfocus lens for the 6-eV laser is placed outside of the UHV main chamber, 300~mm far from the sample.
The beam passes through a quartz vacuum window (the diameter of 36~mm), and the numerical aperture of the beam can be adjusted and optimized to achieve micrometer-level focus at the sample surface.
The photon energy of the ultraviolet lasers is tunable in a range of $h\nu$=5.9-6.49~eV.

In the target chamber, a high-quality Fe(001)$p(1\times$1)-O film is grown on a MgO(001) substrate, and its quality can be assessed using low-energy electron diffraction (LEED).
The prepared target is then transferred $in\; situ$ into VLEED detector of the main chamber.
The cleanliness of the target surface can be maintained for approximately two weeks, but it can be restored by annealing up to 600$^{\circ}$C in the target chamber.

In the MBE chamber, we have installed evaporators, an annealing stage, an ion sputtering gun, LEED, and an auger electron spectroscopy (AES) system.
These systems enable us to prepare atomically controlled thin films $in\; situ$, and confirm the surface order and cleanliness.
\section{Specifications}
\subsection{Spatial resolution}
\begin{figure*}[t]
	\centering
	\includegraphics[width=\linewidth]{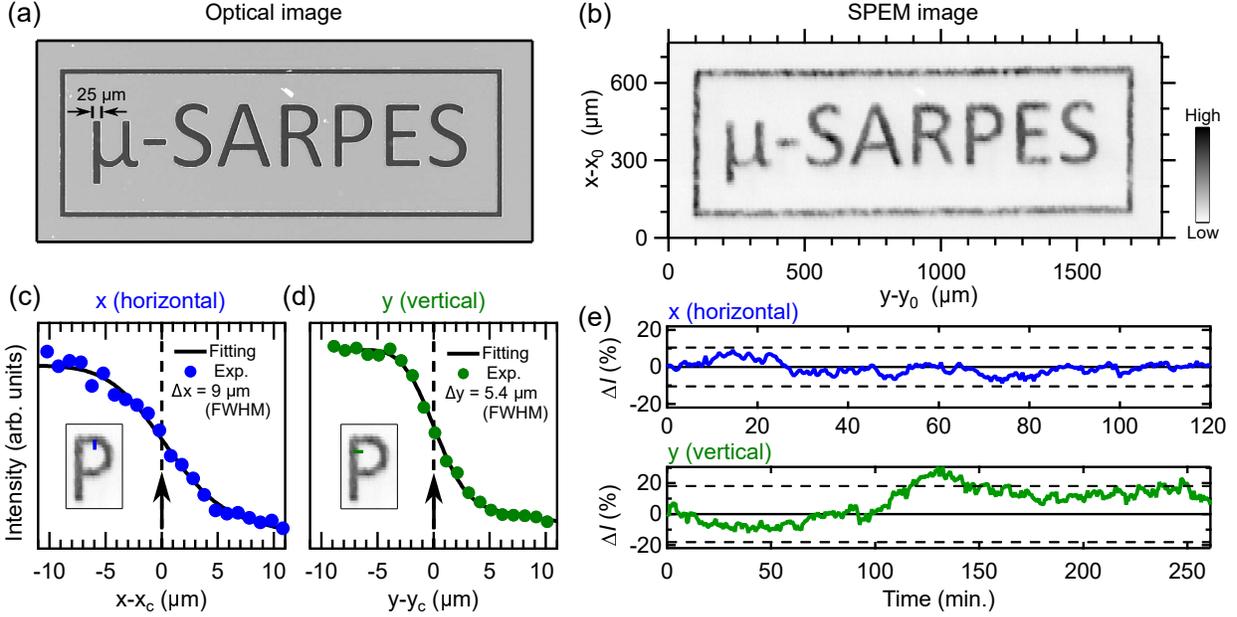}
	\caption{Characterization of the spatial resolution. (a) The optical microscope image of the Au pattern on the Si(001) substrate.
(b) The image of the scanning photoelectron intensity map (SPEM) taken in the Au pattern.
(c) and (d) Photoelectron intensity profiles at the edges of the Au pattern along the $x$ (horizontal) and $y$ (vertical) axes, respectively (green and blue lines in the insets panels).
(e) Plots of the intensity deviation $\Delta{I(x)}$ and $\Delta{I(y)}$ as a function of time to evaluate possible long-term drifts of the laser and the sample stage.
The data was recorded at the intensity edge at $x_c$ and $y_c$ denoted by arrows in (c) and (d).
The 1~{\m}m of the drift along $x$ ($y$) axis can correspond to 19~$\%$ (10~$\%$) of $\Delta{I(x)}\; [\Delta{I(y)}]$ guided by dashed lines in Fig.~\ref{fig2}(e) (see also main text for the details of the estimation). 
}
	\label{fig2}
\end{figure*}

To demonstrate the spatially resolved measurement, we used an Au pattern fabricated by photolithography on a Si(001) substrate [Fig.~\ref{fig2}(a)] in which the width of letters is designed to be 25~{\m}m.
The Au pattern was installed in the UHV chamber and annealed at 200~$^{\circ}$C for several hours to obtain the clean surface before the measurements. We fixed the photon energy of the ultraviolet laser at 6.39~eV (194 nm).

Figure~\ref{fig2}(b) presents a scanning photoemission microscope (SPEM) image~\cite{iwasawa2017elsevier} of the Au pattern.
The SPEM image was obtained by scanning the sample stage with 10~{\m}m and 5~{\m}m step along $x$ and $y$ axes, respectively, and the total pixel number is 27225 points.
The pixel intensity corresponds to the integrated intensity for the emission angle of ($-$15$^{\circ}$, 15$^{\circ}$) and the energy $E-E_{\rm{F}}$ of ($-$0.3~eV, 0~eV). 
It is immediately found that the intensity pattern seen in the SPEM image [Fig.~\ref{fig2}(b)] is very similar to that in the optical microscope image [Fig.~\ref{fig2}(a)].
This dataset already demonstrates the good spatial resolution of our \m-SARPES apparatus.

To further evaluate the spatial resolution quantitatively, we take intensity profiles across the edge of the Au pattern along $x$ and $y$ axes [Figs.~\ref{fig2}(c) and (d), respectively].
By fitting a step function convoluted with a Gaussian function, we estimated the spot size at the sample surface in full width of half maximum (FWHM) to be 9.0~{\m}m and 5.4~{\m}m along the $x$ and $y$ axes, respectively.
The spot size along the $x$ axis is 1.7 times larger than that along the $y$ axis. This is due to the incident angle $\phi$ of the laser on the sample ($\phi$=50$^{\circ}$ in our setup). 
Taking into account the effect of the incident angle, the ratio of the spot sizes along the $x$ and $y$ axes should be approximately 1/cos$\phi$ ($\sim$1.6), which is almost consistent with our results.
 
Besides the spot size, it is very important to consider the long-term drift of the laser and sample stage.
To investigate this stability, we monitored the photoelectron intensity at the edge of the Au pattern [$x_c$, $y_c$ arrows in Figs.~\ref{fig2}(c) and (d)] as a function of time.
In Fig.\ref{fig2}(e), we plot the time evolution of the intensity deviation, defined as $\Delta{I(t)}=[I(t)-I(t=0)]/I(t=0)$.
The drifts along the $x$ and $y$ axes can be estimated by the intensity deviation through $[dI(x)/dx]_{x=x_c}$ and $[dI(y)/dy]_{y=y_c}$ [Figs.\ref{fig2}(c) and (d)].
Hence, 19~$\%$ (10~$\%$) of $\Delta{I(x)}$ [$\Delta{I(y)}$] is considered to correspond to 1{\m}m of drift from the edge [dashed lines in Fig.~\ref{fig2}(e)].
The results guarantee that the drift of the laser and sample stage is suppressed to the sub-micron level, and is much smaller than the spot size of the 6-eV laser, at least for several hours of measurement.
The contribution of laser and stage drifts during the measurement is almost negligible, so the spatial resolution of our \m-SARPES apparatus is mainly determined by the laser spot size.

\subsection{Energy resolution}                                                                                                                                                                                                 
\begin{figure}[t!]
\centering
\includegraphics[width=0.6\columnwidth]{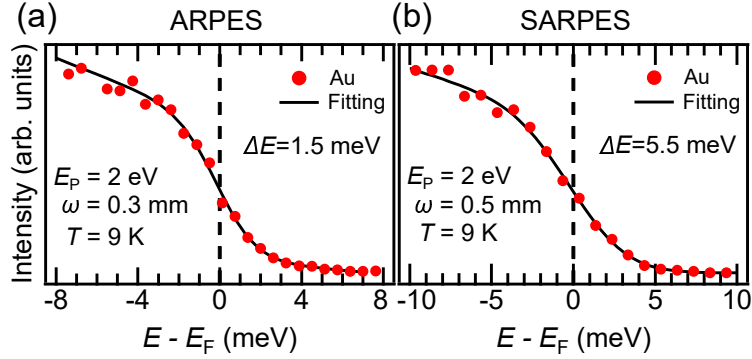}
\caption{
Characterization of the energy resolutions.
(a) ARPES spectrum obtained from an evaporated polycrystalline Au film.
(b) The spectrum obtained by the channeltron of VLEED spin-detector.
Each experimental setting for these measurements is noted in the panels. 
}
\label{fig3}
\end{figure}

We measured polycrystalline Au films deposited on an oxygen-free copper plate to evaluate the energy resolution compatible with the spatial resolution.
Angle-integrated energy distribution curves (EDC) for ARPES are presented in Fig.~\ref{fig3}(a), taken at a temperature ($T$) of 9~K with a pass energy ($E_\text{p}$) of 2~eV and an entrance slit width ($\omega$) of 0.3~mm.
The EDC was fitted with a Fermi-Dirac function convoluted by a Gaussian function (black lines) in Fig.~\ref{fig3}(a).
The total energy resolution ($\Delta E$) is determined to be 1.5 meV, including contributions from the bandwidth of the laser and the electron analyzer setup.
This value is consistent with an estimation based on $\Delta E=E_p\omega/2R=1.5$~meV ($R$ is the radius of the analyzer with $R=200$ mm).

In SARPES mode, the photoelectron count rate for spin detection is typically 10$^{-4}$ times less than that for ARPES.
To compensate for this low efficiency, we increased the analyzer slit up to 0.5~mm and selected an aperture size of 2$\times$1~mm$^2$ for the spin detection.
Under this setup, the energy resolution for SARPES was evaluated to be 5.5~meV.
For practical measurements, a suitable energy (angular) resolution can be chosen in the range of 5.5 meV to 30 meV (0.75~$^{\circ}$ to 1.5~$^{\circ}$) by properly selecting the aperture size.

\section{Performance}
\subsection{Bi$_2$Te$_3$: spin-map and spin-vector detection}
The advantage of using the bright laser for SARPES is that one can quickly map not only electronic band dispersion but also spin polarization.
Here, we demonstrate it through the measurements on a Bi$_2$Te$_3$ single crystal that is known to host a Dirac-cone like surface state on the (111) surface~\cite{YLChen_Sciene2009,hsieh2009nature}.
The clean surface was obtained by cleaving at the UHV environment at room temperature, and SARPES measurements were performed at $T$=30~K. 
We use a Sherman function of 0.30 for the both VLLED-1 and VLEED-2 detectors to obtain spin polarizations.

\begin{figure*}[t!]
\centering
\includegraphics[width=1.0\textwidth]{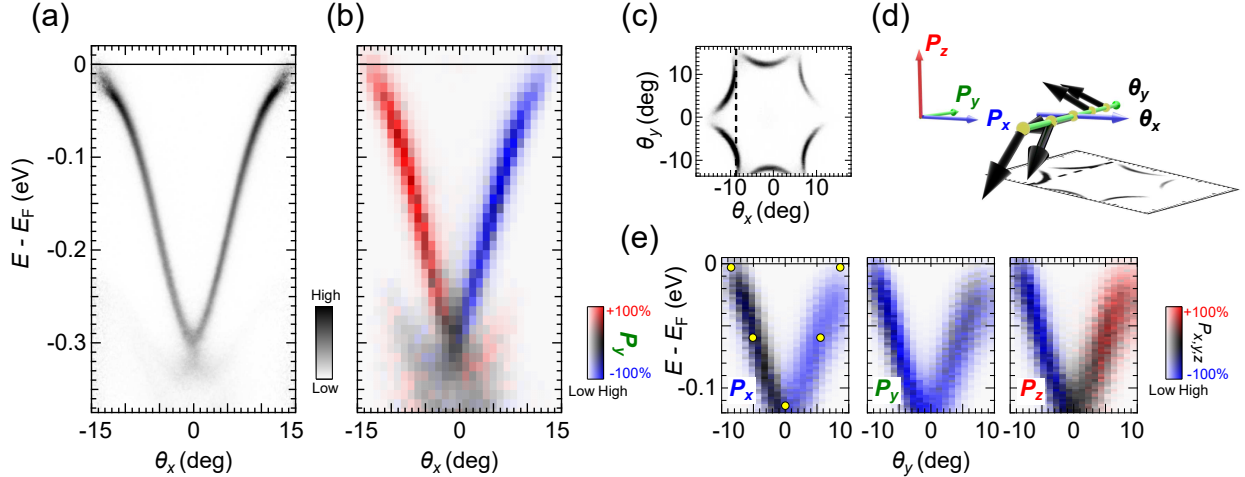}
\caption{Spin-polarization map of Bi$_2$Te$_3$.
(a) ARPES band map acquired at $T$=30~K. 
(b) Spin-polarization map with $P_y$ component, expressed with the two-dimensional color code~\cite{TUSCHE2015ultramicro}.
(c) Fermi surface mapping obtained by ARPES.
(d) The plot of the 3D spin vector determined by our SARPES results. 
(e) Spin-polarization maps for $P_x$, $P_y$, and $P_z$ cut along the angle line denoted by the dashed line in (c). 
The spin polarizations were obtained with the effective Sherman function of 0.30.
}
\label{fig4}
\end{figure*}

Figures~\ref{fig4}(a) and (b) show the high-resolution ARPES band map and the spin-polarization map taken by SARPES, respectively.
In the spin-polarization map [Fig.~\ref{fig4}(b)], the dataset was obtained by sequentially acquiring single spin-resolved EDCs and scanning the emission angle using the deflector function of the electron analyzer ($\theta_x$) without a sample rotation.
The spin-polarized surface state is observed with their expected spin characters quantized along $y$.
For this SARPES measurement, the energy/angular resolution of SARPES was set to be 20~meV/0.75~$^{\circ}$.
Each EDC is obtained in 100 seconds, and the spin polarization map is composed of 31 individual spin-resolved spectra, taken sequentially over a total acquisition time of 50 minutes.
Thus, one can perform the quick spin-polarization map owing to the combination of the 6-eV laser and VLEED spin detector in {\m}-SARPES measurements.

Next, we show the detection of the spin-polarization vector ($\bm{P}$) in three dimensions [$\bm{P}$ is defined as ($P_x$, $P_y$, $P_z$)].
The surface state of Bi$_2$Te$_3$ is known to form a characteristic spin texture in momentum-space where not only the in-plane spin component ($P_x$, $P_y$) but also the out-of-spin-component ($P_z$) appears.
To access the characteristic spin texture, we collected spin-polarization maps along all three axes in Fig.~\ref{fig4}(e) by scanning electron deflector angles along $\theta_y$ [the dashed line in Fig.\ref{fig4}(c)].
From these results, the spin vector $\bm{P}$ can be illustrated by arrows in Fig.~\ref{fig4}(d) at representative emission angles [yellow circles in Fig.~\ref{fig4}(e)].
It is immediately found that our results well capture that the out-of-plane spin component ($P_z$) is reversed for $\pm\theta_y$ while the in-plane components ($P_x$, $P_y$) are insensitive to $\theta_y$.
The observed texture of $\bm{P}$ is consistent with the characteristic feature of the Bi$_2$Te$_3$ surface state~\cite{FuPhysRevLett2009,souma_prl2011}.
\subsection{PbBi$_4$Te$_4$S$_3$: spatially selective measurement}
\begin{figure*}[t!]
	\begin{center}
		\includegraphics[scale=1]{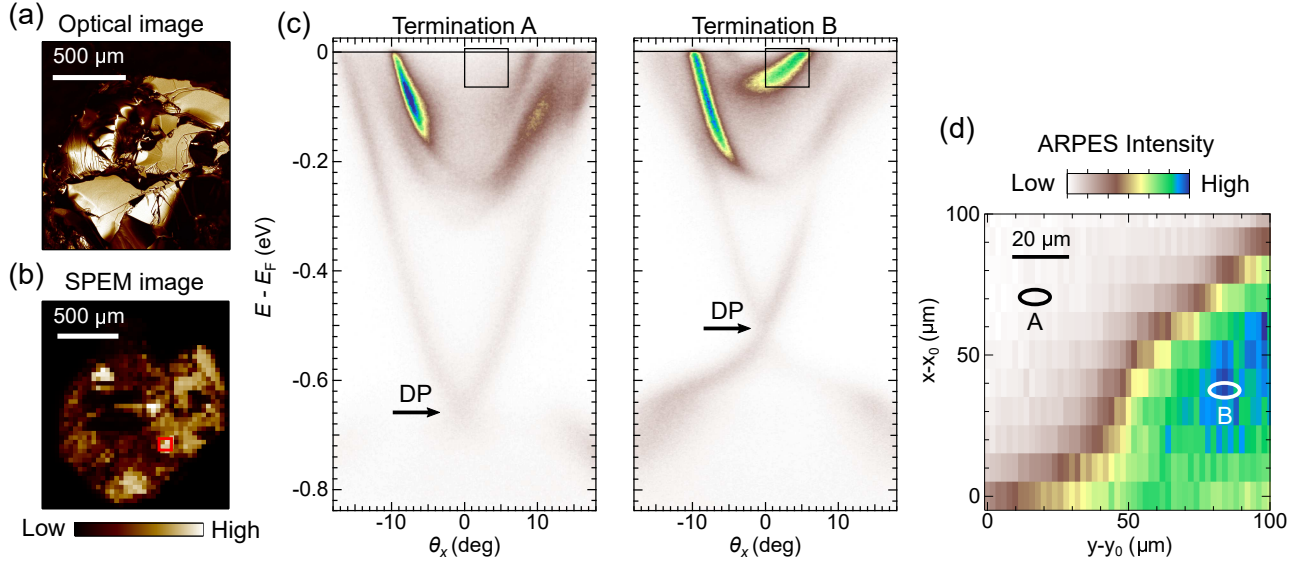}
		\caption{Spatially-resolved measurements on PbBi$_4$Te$_4$S$_3$.
			(a) Optical microscope image of a cleaved surface of PbBi$_4$Te$_4$S$_3$.
			(b) SPEM image taken with a 50~\m m step.
			(c) Spatially-resolved ARPES images acquired at different surface terminations~\cite{sumida2018prm}.
			(d) Scanning ARPES intensity map measured at the 100$\times$100~{\m}m$^2$ marked by the red square in (b).
			The map is taken by scanning 2~{\m}m and 10~{\m}m steps along $x$ and $y$ axes, respectively.
		}
		\label{fig5}
	\end{center}
\end{figure*}
We perform spatial-resolved measurements on another topological insulator PbBi$_4$Te$_4$S$_3$.
In contrast to Bi$_2$Te$_3$ that has a Dirac-cone like surface state as prototype topological insulator, PbBi$_4$Te$_4$S$_3$ has the two surface-states originating from two different terminations~\cite{sumida2018prm}.
Therefore, this material is suitable for demonstrating the measurement with good spatial resolution.
The clean surface was obtained by cleaving at the UHV environment at room temperature, and the measurements are performed at $T$=30~K.

Figure~\ref{fig5}(b) displays a SPEM image of the PbBi$_4$Te$_4$S$_3$ cleaved surface.
The photoelectron intensity in the SPEM image is uneven on a scale of 100 {\m}m, indicating that the sample is not flat and exhibits some bumps and irregularities.
These features are noticeable also in the optical microscope image presented in Fig.~\ref{fig5}(a).

Our measurement with a good spatial resolution allow us to selectively detect the two types of surface states from the distinct surface terminations~\cite{sumida2018prm} (termination A and B).
Accordingly, we obtained ARPES images from the two surface terminations [Fig.~\ref{fig5}(c)]. 
The Dirac-cone like band dispersion is observed in both terminations, but the crossing point, namely Dirac point (DP), is located at different energy; $E-E_{\rm{F}}\sim-0.65$~eV in termination A while $E-E_{\rm{F}}\sim-0.50$~eV in termination B [see arrows in Fig.~\ref{fig5}(c)].

Additional band dispersions are also observed in both terminations within the energy range of $E-E_{\rm{F}}$=0~eV to $-$0.2~eV.
The most striking difference is a strong ARPES intensity observed just below $E_{\rm{F}}$ in termination B.
The ARPES intensity mapped for the two surface terminations results in a microscope image as shown in Fig.~\ref{fig5}(d).
This image was obtained by scanning the sample stage with 10~{\m}m and 5~{\m}m step along $x$ and $y$ axes, while the intensity is integrated in an energy-momentum window shown in Fig.~\ref{fig5}(c).
The different surface terminations are distributed on a scale of approximately several tens of micrometers, and furthermore, their boundary can be clearly observed thanks to the good spatial resolution of our laser {\m}-SARPES apparatus.

Let us note that the spatial map obtained from the angle-resolved intensity [Fig.~\ref{fig5}(d)] represents different information compared to the SPEM obtained from the angle-integrated intensity [Fig.~\ref{fig5}(b)].
The SPEM image similar to an optical microscope reflects the shape of the cleaved surface, such as its flatness and defects.
On the other hand, the angle-resolved map reflects the differences in electronic band structures observed through angle-resolved measurements, as presented in Fig.~\ref{fig5}(d).
Therefore, the spatial map obtained from the angle-resolved intensity not only reveals the surface termination dependence of electronic states, as demonstrated in PbBi$_4$Te$_4$S$_3$ measurement in Fig.~\ref{fig5}, but also proves powerfulness for a wide range of materials, including systems with phase-separated electronic structures.
\section{Summary}                
We have introduced a setup for {\m}-SARPES apparatus with the 6-eV laser and double VLEED spin detectors capable of spin-polarization mapping and spatial-resolved measurements.
Thanks to the high-photon flux laser, the setup achieves an energy resolution of 1.5 (5.5) meV without (with) the spin detection mode, compatible with a spatial resolution better than 10 {\m}m.
The capability of {\m}-SARPES provides a significant advantage in investigating the spatially-resolved spin properties of various quantum materials.

In addition to these advantages of the apparatus, the electron deflection function allows for Fermi surface mapping and spin-polarization mapping without a sample rotation.
This capability provides significant benefits for conducting spatially-resolved measurements, and this advantage extends to measurements on small samples or samples with poor cleavability.
\section{Acknowledgments}
We thank Oleg E. Tereshchenko for providing us single-crystals of Bi$_2$Te$_3$ and PbBi$_4$S$_4$Te$_3$.
We also thank E.~Schwier, K.~Taguchi, T.~Yoshikawa, T.~Warashina, T.~Kono, T.~Sugiyama, Y.~Morita, and R.~Watarizaki for their supports in constructing the machine and conducting the experiments.
This work was supported by the JSPS KAKENHI (Grants Numbers. JP22H01943, JP22H04483, JP21H04652, JP20H00347, JP18H01954, JP16J03874).
%
\bibliographystyle{naturemag}
\bibliography{mSARPES}
\end{document}